\documentclass[figures]{epl}
\usepackage{psfrag}

\title{Vibrational instability due to coherent tunneling of electrons}
\shorttitle{Vibrational instability due to tunneling }

\author{D. Fedorets\thanks{E-mail: \email{dima@fy.chalmers.se}} 
        \and L.Y. Gorelik \and R.I. Shekhter \and M. Jonson }
\shortauthor{D. Fedorets \etal}

\institute{
   Department of Applied Physics, Chalmers University of
  Technology  and G\"oteborg University -
   SE-412 96 G\"oteborg, Sweden
}

\begin{document}

\maketitle

\begin{abstract}
Effects of a coupling between the mechanical vibrations of a 
quantum dot placed between the two leads of a single electron
transistor and coherent tunneling of electrons through  
a single level in the dot has been studied.
We have found that for bias voltages exceeding
a certain critical value a dynamical instability occurs and mechanical
vibrations of the dot develop into a stable limit cycle.
The current-voltage characteristics for such a transistor  were
calculated and they seem to be in a reasonably good agreement with recent 
experimental results for
the single $C_{60}$-molecule transistor by  Park et al.
( Nature {\bf 407,} (2000) 57).
\end{abstract}

\section{Introduction}
Nanoelectromechanics \cite{Roukes00,Craighead00} is a new, quickly
developing field in condensed matter physics.
A coupling between strongly pronounced mesoscopic features of the
electronic degrees of freedom (such as quantum coherence and
quantum correlations) and degrees of freedom connected to
deformations of the material produces strong electromechanical effects
on the nanometer scale. The mesoscopic force oscillations in
nanowires \cite{Rubio96,Iijima96,Yakobson96} 
observed a few years ago is an example of such a phenomenon.
Investigations of artificially-made nanomechanical devices,
where the interplay between single-electron tunneling
and a local mechanical degree of freedom significantly controls the
electronic transport, is another line of nanoelectromechanics
\cite{Tuominen99,Park00,Erbe98,Weiss99,Krommer00,Erbe01,
Gorelik98,Isacsson01, Nord01,Boese00}.
For one of the nanomechanical systems of this kind, the self-assembled
single-electron transistor, a new electromechanical phenomena -
the shuttle instability and 
a new so-called shuttle mechanism of the charge transport 
were recently predicted  in
\cite{Gorelik98}. 
It was  shown that a
small metallic grain attached to two metallic electrodes by elastically
deformable links breaks into oscillations  if a large enough bias voltage is
applied between the leads. 
For the model system studied in \cite{Gorelik98},
 it was also shown that a finite friction is
required for the oscillation amplitude to saturate and 
for a stable regime of oscillations to develop.

An essential assumption made in \cite{Gorelik98} is that the
relaxation mechanisms present are strong enough to keep the electron
systems in
each of the conducting parts of the transistor in local equilibrium (as
assumed in the standard theory of Coulomb blockade
\cite{Shekhter73,Kulik75}).
Such relaxation, which destroys any phase coherence between electron
tunneling events, allows a description of the whole electronic kinetics
 by means of a master equation for the occupation
probabilities of the dot.
It is clear that such an incoherent approach must fail if the size of
the grain is small enough.
Firstly, a decrease of the grain size results in a decrease of its
mass and consequently in
an increase of the frequency of the mechanical vibrations, which for
nanometer-size clusters is of the order of $0.1 \div 1\un{THz}$
and might exceed the electron relaxation rate in the grain.
Secondly, the electron energy level spacing in a nanometer-size grain
$\delta \sim \epsilon_{F}/N$, ($N$ is the number of electrons)  can be
of the order of $10\un{K}$ and exceed the operational
temperatures. 
In this situation the discreteness of the energy spectrum
can be very important.                        
What is more, a  possibly strong tunneling-induced coupling
between electronic states in the grain and in the leads,
which results in large quantum fluctuations
of the charge in the grain, must be included.

For all of the above reasons a new approach is needed for a description
of the electron transport through a nanometer-size movable cluster or
quantum dot. The non-trivial question then arises
whether  or not the coherent electron tunneling through
a movable dot causes any electromechanical instability.
Such a question is of notable practical significance in view of the
recent experiment by Park {\em et al.} \cite{Park00}, where the current
through so-called single-$C_{60}$-molecule transistors was measured and
anomalies --- including a few equidistant step-like features --- in  the I-V
characteristics observed. 
This current steps were interpreted in \cite{Park00} as a
manifestation of the coupling between electron tunneling and
the center-of-mass vibrational degree of freedom of the molecule.

\section{Theoretical model}
\begin{figure}
    \psfrag{x}{x}
    \psfrag{+eV}{$\mu_L=eV$}
    \psfrag{0}{$\mu_R=0$}
    \psfrag{Lead}{Lead}
    \psfrag{Dot}{Dot}
    \onefigure[width = 7cm]{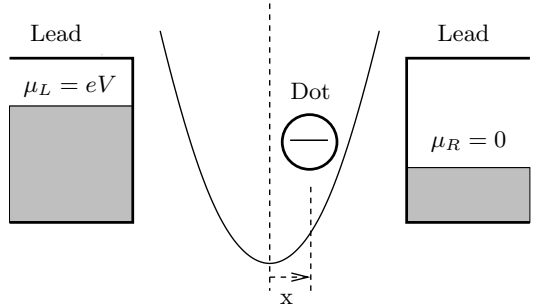}
    \caption{Model system consisting of a movable quantum dot placed
    between two leads.
    An effective elastic force acting on the dot from the leads is
    described by the parabolic potential.
    Only one single electron state is available in the dot
    and the non-interacting electrons in each have a
    constant density of states.
      }\label{system}
  \end{figure}

To investigate the influence of the above electromechanical coupling
we consider a model system consisting of a movable quantum dot placed
between two bulk leads.
An effective elastic force acting on the dot due to interaction with
the leads is described by the parabolic potential 
presented in fig.~\ref{system}.
We assume that only one single electron state is
available in the dot and that the electrons in each lead are
non-interacting with a  constant density of states. At the same time we 
treat the motion of the  grain classically. The Hamiltonian for the electronic
part of
the system is
\begin{equation}
H= \sum_{\alpha, k} ( \epsilon_{\alpha k}- \mu_\alpha)
a_{\alpha k}^{\dag} a_{\alpha k}+ \epsilon_d(t)c^{\dag} c+
\sum_{\alpha, k} T_{\alpha}(t)
(a_{\alpha k}^{\dag} c + c^{\dag} a_{\alpha k})\,.
\end{equation}
Here $T_{L, R} =  \tau_{L, R} \exp\{\mp x(t)/\lambda\}$ is the
position-dependent tunneling matrix element,
$\epsilon_d(t) = \epsilon_0-{\cal E}x(t)$
is the energy level in the dot shifted due to the
voltage induced electric field ${\cal E}/e = \chi V$, 
$\chi$ is a parameter characterizing the strength of the electrical field
as a function of the bias voltage $V$, 
$e<0$ is the electron charge, 
$x$ measures the displacement of the dot,
$a_{\alpha k}^{\dag}$ creates an electron with momentum $k$ in
the corresponding lead, $\alpha = L, R$ is the lead index,
$c^{\dag}$ creates  an electron in the dot and 
$\lambda$ is the characteristic tunneling length \cite{Gorelik98}.
The first term in the Hamiltonian describes the electrons in the leads,
the second  --- the movable quantum dot 
and the last term --- tunneling between the leads and the dot.

The evolution of the electronic subsystem is determined by the
Liouville-von
Neumann equation for the statistical operator $\hat{\rho}(t)$,
\begin{equation}
  \label{neumann}
  i\partial_t \hat{\rho}(t)= [\hat{H},\hat{\rho}(t)]_{-} \,,
\end{equation}
while the center of mass motion of the dot is governed by Newton's
equation,
\begin{equation}
\label{newton}
\ddot{x}  + w_0^2 x =  F/M\,.
\end{equation}
Here $w_0=\sqrt{k/M}$, $M$ is the mass of the grain,
$k$ is a constant characterizing the strength of the harmonic potential,
$F = -<\partial \hat{H}/ \partial x>$ and
$<\bullet> = Tr\{\hat{\rho}(t)\bullet\}$.
The force $F$ on the RHS of eq.~(\ref{newton}) is due to the coupling
to the electronic subsystem and consists of two terms,
\begin{equation}
F(t) = -i{\cal E} G^{<}(t,t) + 2 \lambda^{-1} \sum_{\alpha, k}(-1)^{\alpha}
T_{\alpha}(t){\rm Im}\left[G_{\alpha k}^{<}(t,t)\right] \,,
\label{forceDef}
\end{equation}
where $G^{<}(t,t') \equiv i\langle c^{\dag}(t') c(t)\rangle$, 
$G_{\alpha k}^{<}(t,t') \equiv i\langle
a_{\alpha k}^{\dag}(t')c(t)\rangle$  are the lesser Green functions,
and  $\alpha=0\,(1)$ for the left (right) lead.
The first term describes the electric force that acts on the charge in
the dot; the second term is an exchange force, which appears due to
the position dependence of the tunneling matrix elements $T_{L,R}$.
The force $F$ depends on the correlation functions
$G^{<}(t,t)$ and $G_{\alpha k}^{<}(t,t)$,
which can be computed exactly in the wide-band limit ($\rho_\alpha=const$) 
by using the Keldysh formalism \cite{Keldysh65}.

Following the standard analysis  \cite{Wingreen94} we express the
correlation function $G_{\alpha k}^{<}(t,t)$ in terms of the Green
functions of the dot,
\begin{equation}
G_{\alpha k}^{<}(t, t') =
\int \, \upd t_{1}  T_{\alpha}(t_{1}) \left\{G^{r}(t,t_{1})
g_{\alpha k}^{<}(t_{1},t') + G^{<}(t,t_{1})
g_{\alpha k}^{a}(t_{1},t')\right\}\,.
\end{equation}
Here  $G^{r}$ is the retarded  Green function of the dot
and $g_{\alpha k}^{a(<)}$ is the advanced (lesser) Green function of
the leads for the uncoupled system.
The Dyson equation for the retarded (advanced) Green function has the
following form:
$G^{r(a)}=g^{r(a)} + g^{r(a)} \Sigma^{r(a)} G^{r(a)}$,
where 
$\Sigma ^{r(a)}(t_{1},t_{2}) = \sum_{\alpha, k}
 T_{\alpha}(t_{1}) g^{r(a)}_{\alpha, k}(t_{1},t_{2})
 T_{\alpha}(t_{2})$.
In the wide-band limit 
$\Sigma ^{r(a)}(t_{1},t_{2}) \propto \delta(t_{1}-t_{2})$ and this
Dyson equation can be solved exactly.
The lesser Green function $G^<$ is given by the Keldysh equation 
$G^< = G^r \Sigma^< G^a$,
where
$\Sigma ^{<}(t_{1},t_{2}) = \sum_{\alpha, k}
 T_{\alpha}(t_{1}) g^{<}_{\alpha k}(t_{1},t_{2})
 T_{\alpha}(t_{2})$.

As a result we obtain a general expression for the force $F$ of the form
\begin{eqnarray}
F(t) =  \sum_{\alpha} \rho_\alpha \int \,
\upd \epsilon f_{\alpha}(\epsilon) \left\{
{\cal E}\left| B_{\alpha}(\epsilon,t) \right|^{2}
+2\frac{(-1)^{\alpha}}{\lambda}  T_{\alpha}(t)  {\rm Re}
[B_{\alpha}(\epsilon,t)] \right\}  \,,
\label{force}
\end{eqnarray}
where
\begin{displaymath}
B_{\alpha}(\epsilon,t) = -i\int_{-\infty}^{t} \,\upd t_1  T_{\alpha}(t_1)
 \exp \left\{i\int_{t_1}^{t}\,\upd t_2 \left[\epsilon-\epsilon_d(t_2)+
i\frac{\Gamma(t_2)}{2}\right]\right\}\,,
\end{displaymath}
$\Gamma(t)= 2\pi\sum_\alpha \rho_\alpha T_\alpha^2(t)$,
$\rho_\alpha$ is the density of states in the
corresponding  lead  and
$f_\alpha(\epsilon) = [\exp(\beta(\epsilon-\mu_\alpha))+1]^{-1}$.
It is worth mentioning that eq.~(\ref{force}) is 
valid for arbitrary values of the tunneling matrix elements. 

An important question is now whether or not the mechanical stability of
the transistor configuration is affected by coherently tunneling electrons.

\section{Instability}
In order to investigate the stability of the equilibrium
position of the dot we expand  the RHS of eq.~(\ref{newton})
to first order with respect to the displacement:  
$F(t) = \int_{-\infty}^{t}\,\upd t' D(t-t')x(t')$. After a  Fourier
transformation of the obtained equation we get the following dispersion 
equation for the frequency:
\begin{equation}
w^2-(w_0^2-w_1^2)=-D_w/M\,,
\end{equation} 
where 
\begin{displaymath}
 D_{w} = - \int \frac{\upd \epsilon}{2\pi} 
\sum_{\alpha} \Gamma_\alpha  f_{\alpha}
\{R_\alpha G_{w}^{+} + R_\alpha^{\ast}G_{-w}^{-}\}\,,
\end{displaymath}
\begin{displaymath}
R_\alpha = 
\left\{ \left|{\cal E} G^{+}_0+\frac{(-1)^{\alpha}}{\lambda}\right|^2 -
ig
\left[{\cal E}  \left|G_0\right|^2
+G^{+}_0 \frac{(-1)^{\alpha}}{\lambda} \right]\right\}\,,
\end{displaymath}
$w_{1}^{2} = \sum_\alpha \Gamma_\alpha \int \,\upd \epsilon 
f_\alpha{\rm Re}[G_0^{+}]/(\pi \lambda^2 M)$,
$\Gamma_\alpha = 2 \pi \rho_\alpha \tau_\alpha^2$,
$\Gamma  = \sum_\alpha \Gamma_\alpha$,
$G^{\pm}_{w}(\epsilon) = [\epsilon + w-\epsilon_{0}
\pm i\Gamma/2 ]^{-1}$ and 
$g=(\Gamma_L-\Gamma_R)/\lambda$.
 
The criterion for instability is that the frequency $\omega$ has a
positive imaginary part.
When $\Gamma/(Mw_0^2 \lambda^2) \ll 1\,$ and 
${\cal E}\lambda / (\sqrt{w_0^2+\Gamma^2}) \ll 1\,$
(the case of  weak electromechanical coupling)
we obtain that ${\rm Im}[w] \approx
-{\rm Im}[D_{w_0}]/(Mw_0)\,$.
From now on we consider only the case of  weak electromechanical coupling.
An analytical analysis can not be performed
 in the general case, but for a large symmetrically
applied bias voltage and zero temperature one can show that
${\rm Im}[D_{w_0}$] is negative:
\begin{equation}
\label{asymptote}
{\rm Im} [D_{w_0}] \asymp -\frac{4{\cal E} w_0}
{\lambda(w_0^{2}+\Gamma^2)} \frac{\Gamma_L \Gamma_R}{\Gamma}\,.
\end{equation}

The instability that follows from eq.~(\ref{asymptote}) is in
contrast with the behavior at low voltages where it can be shown that
${\rm Im}[D_{w_0}$] is positive.
Therefore, a finite threshold voltage for the instability exists in the
system.
A simple expression for the threshold voltage  can be found for a
symmetric junction in the case
of weak tunneling under the above conditions:  $eV_c = 2(\epsilon_0 +
w_0)$.

\begin{figure}
    \onefigure[width = 8cm]{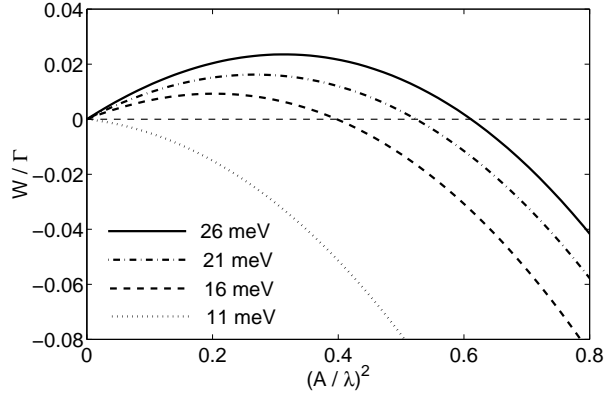}
    \caption{The energy $W$ pumped into the mechanical degree of freedom
      during one period of oscillation  for different values of the
      applied voltage $V$:
           $w_0 = 5\un{meV}$, $ T =  0.13\un{meV}$,
           $ \epsilon_0 = 6\un{meV}$, $\Gamma = 1\un{\mu eV}$
            and $\Gamma_R  /  \Gamma_L = 9$.} \label{energy}
\end{figure}

Under the condition of  weak electromechanical
coupling, the displacement $x(t)$ of the dot can be represented in a
 harmonic oscillation form $A(t)cos(wt+\phi(t))$ with an 
amplitude $A(t)$ and a phase $\phi(t)$ that slowly vary on the scale of
$w_0^{-1}$.
By  averaging over the fast harmonic oscillations one can get the
following equation for the evolution of the amplitude 
(see for example  \cite{Gorelik98}):
\begin{equation}
\frac{dA^2}{dt} = \frac{W(A^2)}{M \pi w_0}\,.
\label{dA^2}
\end{equation}
Here $W(A^2)=\oint \,\upd x  F$ is the
work done by the force $F$ on the dot  during one
period of oscillation with a constant amplitude $A$.
We conclude from eq.~(\ref{dA^2}) that the stable regime of oscillations
corresponds to $W=0$ and $dW(A^2)/dA^2<0$. 
 Typical $W(A^2)$-curves are
depicted in fig~\ref{energy} for the parameters taken from the
experiment described in \cite{Park00}.
When an applied voltage is lower than the threshold voltage 
the work $W$ is negative for all amplitudes.
This implies that the equilibrium position of the grain is stable.
When the voltage is higher than the threshold voltage the function
$W(A^2)$  is positive for $0 < A < A_c(V)$ 
(which corresponds to a slow increase of the oscillation amplitude), 
negative for $A > A_c(V)$ 
(which corresponds to a slow decrease of the amplitude) and equal to
zero at $A = A_c(V)$.
This  means that when the applied voltage exceeds the
threshold value the amplitude of the oscillation slowly 
increases until it develops into a stable limit cycle with amplitude 
$A_c(V)$.
As one can see from fig.~\ref{energy}, the amplitude of the limit
cycle is of the order of $\lambda \sim 0.1\un{nm}$ and
is considerably larger than the amplitude of zero-point oscillations of
the grain, which for the $C_{60}$ molecule in the experiment
\cite{Park00}
is approximately $3\un{pm}$.

\section{Current}
In the regime where the stable limit cycle oscillations have 
developed  the vibration-induced inelastic  
tunneling of electrons can significantly contribute to the current.
The time-averaged current through the system in the stable regime where
oscillations of the limit cycle amplitude $A_c(V)$ and 
frequency $w_0=\sqrt{k/M}$ have developed has the following form:
\begin{eqnarray}
\label{currentGeneral}
 I = -\frac{e\Gamma_{L}}{2\pi T} 
\int_0^{T} \,\upd t \int \,\upd \epsilon \left\{ e^{-\frac{2x}{\lambda}}
\sum_{\alpha} \Gamma_{\alpha} 
f_{\alpha}\left|\frac{B_{\alpha}}{\tau_{\alpha}}\right|^{2}
 +2 e^{- \frac{x}{\lambda}}f_{L}
{\rm Im}\left[ \frac{B_{L}}{\tau_{L}} \right] \right\}\,,
\end{eqnarray}
where $x(t) = A_c cos(w_0 t)$ and 
$T = 2\pi/w_0$ is the period of oscillations. 
This expression is valid under the same general conditions
as eq.~(\ref{force}).

\begin{figure}
    \onefigure[width = 8cm]{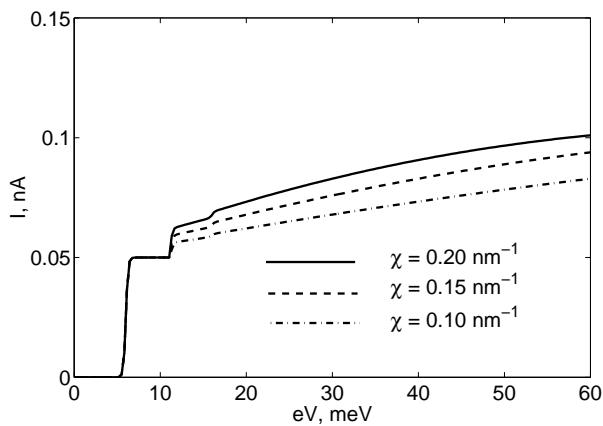}
    \caption{I-V curves for different values of the parameter
    $\chi$ which characterizes the strength of the
     electric field between the leads for a given voltage:
     $w_0 = 5\un{meV}$, $ T =  0.13\un{meV}$,
     $ \epsilon_0 = 6\un{meV}$, and $\Gamma=2.3\un{\mu eV}$. 
     Best fit to \cite{Park00} 
     is obtained for an asymmetric coupling to the leads; here we use 
     $\Gamma_R  / \Gamma_L = 9$.
    }
   \label{I-V}
\end{figure}

Under the condition that the effective level broadening
$\tilde{\Gamma} \ll w_0$ 
(where $\tilde{\Gamma} = \Gamma J_0(i2A_c/\lambda)$) 
eq.~(\ref{currentGeneral}) can be simplified to 
\begin{eqnarray}
\label{current}
I \approx \frac{e\Gamma_L \Gamma_R}{\Gamma}
\sum_{m=-\infty}^{+\infty}
\{ f_{L,m} \xi_{L}^{2m} -
f_{R,m} \xi_{R}^{2m} \} J_m^2(A_c \eta) \,,
\end{eqnarray}
where $f_{\alpha,m} = f_\alpha(\epsilon_0+mw_0)$,
$\eta = \sqrt{({\cal E}/w_0)^2-\lambda^{-2}}$,
$\xi_\alpha = -[ ({\cal E}/w_0)+(-1)^{\alpha}/{\lambda} ] / \eta$
and $J_m$ are Bessel functions of the first kind.
The typical I-V curves are shown in fig.~\ref{I-V},
where we assume that voltage has been applied only to the left lead (according
to the experiment \cite{Park00}). 
The main characteristic feature of all obtained I-V curves is that
they show only a few equidistant steps which are followed  by step-less
behavior of the curves. 
The distance between the steps is given by the vibrational frequency $w_0$
and their heights can vary depending  on the parameters.
The steps following the first one is due to the development of a
vibrational instability and  a transition into 
the associated charge transfer regime.
The obtained behavior of the I-V-curves is in reasonably 
good agreement with the experimental data \cite{Park00}. 
Best fit to the published experimental I-V-curves
is obtained for an asymmetric coupling to
the leads ($\Gamma_R  / \Gamma_L \approx 9$). 
When  the ratio $\Gamma_R  / \Gamma_L $ decreases both the vibration-induced
current jumps and a high-voltage slope of the I-V-curve increase
deviating from the experimental data.

An alternative theoretical description of the 
experiment based on a photon-assisted tunneling-like picture 
\cite{Boese00} also shows reasonable agreement with the experimental data 
\cite{Park00}.
Therefore it is difficult to conclude with certainty that it is
the above instability which is responsible for the specific features
observed in the experiment. Further experiments are needed to clarify
the picture.
Such experiments may involve an investigation of the charge
fluctuation on the grain or of current noise.

\section{Conclusion}
We have studied the effect of a coupling between
coherent tunneling of electrons through a single quantized energy level
in the central island --- or dot --- of a single electron transistor and
vibrations of the dot.  
We have found that for bias voltages exceeding a certain
critical value a dynamical instability occurs and mechanical vibrations of the
center of mass of the dot develop into a stable limit cycle.
The effect of this vibrations on the current through the system were
also studied.
I-V characteristics calculated in our model were found to
be in a reasonably good agreement with recent experimental 
results of Park {\em et al.}\cite{Park00} 
for the single $C_{60}$-molecule transistor.

\acknowledgments
Financial support from the Swedish SSF through the
QDNS program (D. F., L. G.) and the Swedish NFR/VR (R. S.) 
and the U.S. Department of Energy
Office of Science through contract No. W-31-109-ENG-38 
is gratefully
acknowledged.

\end{document}